\documentclass[12pt]{article}

\usepackage{scicite}

\usepackage{times}

\usepackage{amsmath}
\usepackage{amssymb}
\newcommand{\mycite}[1]{\cite{#1}}
\usepackage{siunitx}
\usepackage{amsthm}
\newcommand{\f}{f(t)}
\newcommand{\fsq}{f^2(t)}
\newcommand{\Gammar}{2\pi \times\SI{94(4)}{\kilo\hertz}}
\newcommand{\Gammab}{2\pi\times\SI{82(3)}{\kilo\hertz}}
\newcommand{\nb}{0.8(1)}
\newcommand{\nr}{0.89(8)}
\newcommand{\naddone}{1.4(2)}
\newcommand{\naddtwo}{2.77(6)}

\newenvironment{sciabstract}{%
	\begin{quote} \bf}
	{\end{quote}}

\usepackage{upgreek}
\newcommand{\micro}{\upmu}
\newcommand{\second}{\textrm{s}}
\newcommand{\hertz}{\textrm{Hz}}
\newcommand{\mega}{\textrm{M}}
\newcommand{\giga}{\textrm{G}}

	\usepackage{calrsfs} 
\DeclareMathAlphabet{\pazocal}{OMS}{zplm}{m}{n}
\newcommand{\cC}{\pazocal{C}}
\newcommand{\onehalf}{\frac{1}{2}}
\usepackage{graphicx}
\newcommand{\fc}{f_{\scriptsize\textrm{c}}}

\newcommand{\numeas}{\nu_{\scriptsize\textrm{meas}}}
\newcommand{\numech}{\nu}
\newcommand{\fmone}{f_{\scriptsize\textrm{m},1}}
\newcommand{\fmtwo}{f_{\scriptsize\textrm{m},2}}

\newcommand{\reportednumeas}{{0.44^{+0.004}_{-0.004}~ (\textrm{stat})^{+0.022}_{-0.021}~(\textrm{sys})}}

\newcommand{\reportednumech}{{0.18^{+0.03}_{-0.02}~(\textrm{stat})^{+0.13}_{-0.11}~(\textrm{sys})}}
\newcommand{\reportednumechdB}{{4.43^{+0.6}_{-0.7}~(\textrm{stat})^{+4.4}_{-2.3}~(\textrm{sys})}}
\newcommand{\etaone}{0.26(2)}
\newcommand{\etatwo}{0.153(3)}
\newcommand{\nonecold}{0.75(1)}
\newcommand{\ntwocold}{0.63(1)}
\newcommand{\nonehot}{10.9(1)}
\newcommand{\ntwohot}{4.63(5)}
\newcommand{\Vone}{V_1}
\newcommand{\Vtwo}{V_2}

\newcounter{parnum}
\newcommand{\N}{%
	\noindent\refstepcounter{parnum}%
	\hspace{-1cm}
	\makebox[\parindent][l]{\textbf{\# \arabic{parnum}}}}
\renewcommand{\N}{}

\title{Direct observation of deterministic macroscopic entanglement} 

\author
{Shlomi Kotler$^{1,2}$\footnote{Present address: Department of Applied Physics, The Hebrew University of Jerusalem, Jerusalem, 9190401, Israel.}, Gabriel A. Peterson$^{1,2}$, Ezad Shojaee$^{1,2}$, \\ 
	Florent Lecocq$^{1,2}$, Katarina Cicak$^1$, Alex Kwiatkowski$^{1,2}$,\\ 
	Shawn Geller$^{1,2}$, Scott Glancy$^1$, Emanuel Knill$^{1,3}$,\\
	Raymond W. Simmonds$^1$, Jos\'e Aumentado$^1$, \& John D. Teufel$^1$\\
	\\
	\normalsize{$^{1}$National Institute of Standards and Technology, Boulder, CO 80305, USA.}\\
	\normalsize{$^{2}$Department of Physics, University of Colorado, Boulder, CO 80309, USA.}\\
	\normalsize{$^{3}$Center for Theory of Quantum Matter, University of Colorado, Boulder, CO 80309, USA.}\\
	\\
	\normalsize{$^\ast$To whom correspondence should be addressed; E-mail:  shlomi.kotler@mail.huji.ac.il}
}

\date{}

\begin{document} 
	
	
	\baselineskip24pt
	
	
	\maketitle

	
\begin{sciabstract}
	Quantum entanglement of mechanical systems emerges when distinct objects move with such a high degree of correlation that they can no longer be described separately. Although quantum mechanics presumably applies to objects of all sizes, directly observing entanglement becomes challenging as masses increase, requiring measurement and control with a vanishingly small error. Here, using pulsed electromechanics, we deterministically entangle two mechanical drumheads with masses of 70 pg. Through nearly quantum-limited measurements of the position and momentum quadratures of both drums, we perform quantum state tomography and thereby directly observe entanglement. Such entangled macroscopic systems are uniquely poised to serve in fundamental tests of quantum mechanics,  enable sensing beyond the standard quantum limit, and function as long-lived nodes of future quantum networks. 
\end{sciabstract}

\N
The idea that motion has a non-classical nature dates back to the early days of quantum mechanics. One of the first triumphs of the theory was explaining the emission and absorption spectra of atoms by quantizing the motion of their electrons. Quantum mechanics is not limited to the atomic scale; in principle it extends to all objects of all sizes. We expect that quantum behavior of macroscopic systems will enhance our ability to build more powerful sensing, communication, processing and storage devices~\cite{Safavi-Naeini:2019}.  


\N
Many future applications of quantum technology rely heavily on entanglement, that is, on the ability to generate strong quantum correlations between separate objects. For entanglement to be useful, it must be prepared efficiently, followed by measurement and control with precision that is inversely proportional to the square root of the masses of the objects involved. The task becomes more difficult in the presence of noise, especially given the facts that larger objects tend to interact more strongly with noisy environments and that the measurement process also introduces noise (see Fig.~1A). The communication or processing protocol in which the entanglement might be used limits the amount of noise allowed before the entanglement is rendered useless. 

\N
Entanglement of mechanical motion was demonstrated for the first time with two trapped atomic ions~\cite{Jost2009}.  It was generated deterministically and measured directly with high fidelity, and was therefore available as a resource that could be used for further processing. Taking the same level of quantum control and measurement from the atomic scale to macroscopic engineered objects then remained an outstanding challenge. Important experimental milestones towards this goal have been reached either using optical photons in a probabilistic scheme~\cite{Riedinger:2018hw} or using microwave radiation with indirect inference~\cite{Ockeloen-Korppi2018}.

\N
Here we strongly and deterministically entangle two massive mechanical oscillators and directly observe their state. Our technology allows for on-demand reproducible entanglement generation. For direct state observation, we implement a near quantum-limited measurement of the position and momentum quadratures of both mechanical oscillators in every realization of the experiment. By repeating these measurements, we completely characterize their joint covariance matrix. This tomography demonstrates clear evidence of continuous variables (CV) entanglement~\cite{RMPBraunstein2005} in the measurement signals, without noise subtraction. 

\N
The entangled state measured here manifests strong correlations between seemingly disparate systems. The two-oscillator system can be characterized by the first and second moments of the dimensionless quadratures of motion $X_j$ and $P_j$ ($j=1,2$) which satisfy the canonical commutation relations $[X_j,P_j]=i$, and are related to the underlying mechanical positions and momenta~\cite{Caves:1980}. After entanglement generation, the $X$-quadrature of each harmonic oscillator is drawn from a  Gaussian probability distribution with a variance that is large compared to its zero-point fluctuations. However, when compared against each other, $X_1$ and $X_2$ are highly correlated, and similarly $P_1$ and $P_2$ are anti-correlated. These features, however striking, could be consistent with classical correlations. To verify that the correlations originate from entanglement, we use the Simon-Duan criterion~\cite{Simon2000,Duan2000,BowenBook}, calculated from the covariance matrix $\cC$ of $\vec{S}=\left(X_1,P_1,X_2,P_2\right)$. Element $j,k$ of the $4\times 4$ covariance matrix is $\cC_{jk}=\onehalf\langle (S_k-\langle S_k\rangle)(S_j-\langle S_j\rangle)+(S_j-\langle S_j\rangle)(S_k-\langle S_k\rangle)\rangle$ where $\langle\ldots\rangle$ denotes expectation value. The smallest symplectic eigenvalue $\numech$ of the partially-transposed covariance matrix, quantifies the entanglement of the system~\cite{Serafini_2003,BowenBook} (see~\cite{Kotler:2020SI} for explicit expressions). The two-oscillator state is entangled if $\numech< \onehalf$, where the zero-point fluctuations have variance $\onehalf$.  

\N
To extract the full covariance matrix with minimal assumptions, we measure all four quadratures of motion describing the two oscillators in each single experiment. Our method is analogous to a heterodyne measurement of electromagnetic radiation and allows more accurate covariance matrix estimation than its homodyne counterpart~\cite{Teo2017}. Crucially, this concurrent measurement improves substantially if it is efficient. The inefficiency of our measurement apparatus can be modeled as an effective beam-splitter~\cite{Palomaki2013,Palomaki710,Delaney2019}, where each variable describing the motion $S_j\in\left\{X_1,P_1,X_2,P_2\right\}$, becomes mixed with vacuum noise such that, $s_{j}=\sqrt{\eta_j}S_j +\sqrt{1-\eta_j}\xi_j$, where $\eta_j$ is the efficiency of the measurement of oscillator $j$, $\xi_j$ is a Gaussian random variable with zero mean and vacuum variance $\langle \xi_j^2\rangle=\onehalf$, and $\xi_j$ and $\xi_k$ are independently distributed for $j\neq k$. Therefore, following calibration of the measurement chain, we have direct access to the measured variables $s_{j}$ and their 
 minimal symplectic eigenvalue $\numeas$. For the states measured here, $\numeas<\onehalf$ implies that $\numech<\onehalf$ and vice versa~\cite{Kotler:2020SI}. This mutual relation takes an even simpler form if in addition the state is symmetric with respect to exchanging the roles of the two oscillators and undergoes symmetric loss: $\numeas=\eta \numech+(1-\eta)\onehalf$. In both cases, if efficiencies are low, $\numeas$ will approach $\onehalf$, and it will be more difficult to certify that $\numech<\onehalf$ with high confidence. Moreover, quantum information protocols, such as teleportation and entanglement swapping, require a high measurement efficiency, as demonstrated for light fields~\cite{Furusawa706,Jia2004}.

\begin{figure} 
	\centering
	\includegraphics[scale=1.0]{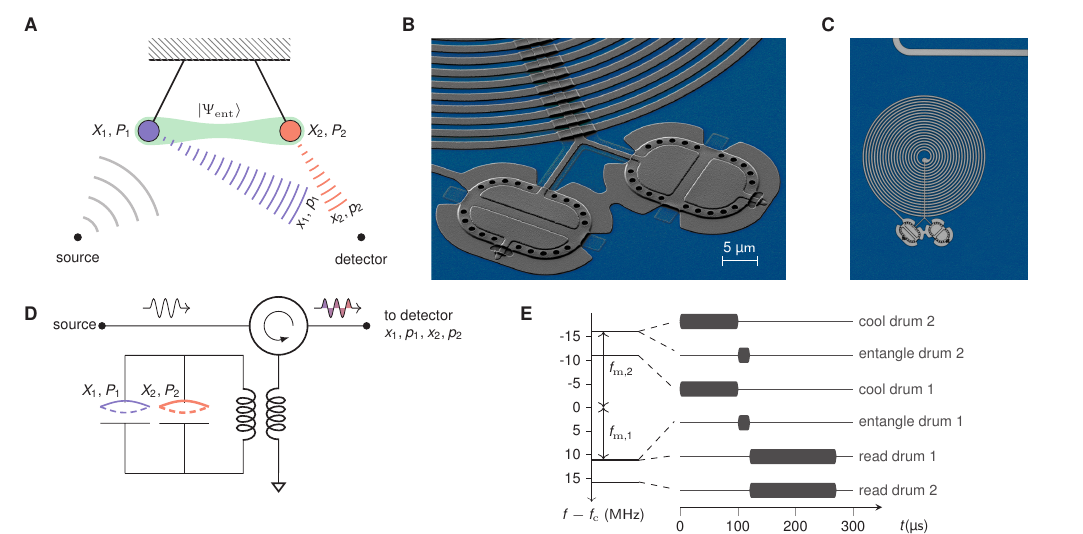}
	\caption{
		{\bf Experiment overview.}
		\textbf{(A)} Concept. Two mechanical harmonic oscillators (pendulums), characterized by their respective quadratures of motion $X_1,P_1,X_2,P_2$, are placed in an entangled state $| \psi_{\scriptsize\textrm{ent}}\rangle$. Electromagnetic radiation Doppler-shifts as it reflects off the moving pendulums, carrying information mixed with noise: $x_1,p_1,x_2,p_2$, due to inevitable loss effects.  
		\textbf{(B)} Scanning electron micrograph (false color) of a device similar to the one used in this paper. Two aluminum drums are suspended above a sapphire substrate, resulting in well-defined harmonic modes in the direction perpendicular to the substrate with frequencies $\fmone=\SI{10.9}{\mega\hertz}$ (left drum) and $\fmtwo=\SI{15.9}{\mega\hertz}$ (right drum). Each drum forms the top plate of a capacitor, along with the bottom plate which is fixed to the substrate.
		\textbf{(C)} Device optical image (false color). A spiral inductor shunts the parallel capacitance of the two drums. Together they form a microwave cavity at a frequency of $\fc=\SI{6.0806}{\giga\hertz}$. An input line (top right) inductively couples to the microwave cavity.
		\textbf{(D)} Circuit schematics. Mechanical motion (dashed) modulates the frequency of the microwave cavity. Therefore, an incoming pulse is Doppler-shifted as it reflects off the cavity, encoding information about the drums' quadratures of motion. Incoming and reflected pulses are separated using a circulator. 
		\textbf{(E)} Experiment sequence. The carrier frequency of the incoming pulses determines the nature of their interaction with the drums. State initialization is achieved by sideband cooling each drum mode close to its ground state followed by a short entangling pulse.  A readout pulse imprints an amplified record of the mechanical states onto the reflected microwave pulse. 
	}
\end{figure}

\N
Our two mechanical oscillators are made of lithographically-patterned thin-film aluminum that forms drum-like membranes~\cite{Teufel2011}, each with a mass of $\approx 70~\textrm{pg}$, suspended above a sapphire substrate (Fig.~1B). We use the $\fmone=10.9$ MHz mode of the left drum and the $\fmtwo=15.9$ MHz mode of the right drum. We manipulate and measure the motion of the drums using electromechanics~\cite{BowenBook}. The drums are embedded into a single microwave resonator, known as the `cavity', whose resonance frequency, centered at $\fc=6.0806$ GHz, shifts according to the drums' motion (Fig.~1C, 1D). A microwave pulse, reflected off the cavity, imparts forces on the drums and encodes the amplitudes of their quadratures of motion into Doppler-shifted sidebands of the microwave pulse. The carrier frequency of the incoming microwave pulse determines whether the drums are cooled, entangled or measured (Fig.~1E). 

\N  
To entangle the two drums, we irradiate the cavity with two pulses simultaneously. One pulse has a carrier frequency of $\fc+\fmone$. This results in a two-mode squeezing (TMS) interaction that entangles the cavity with drum 1, by generating correlated photon-phonon pairs: photons at a frequency of $\fc$ and phonons at $\fmone$~\cite{Palomaki710}. The other pulse has a carrier frequency of $\fc-\fmtwo$. This results in a beam-splitter (BS) interaction that swaps cavity photons at a frequency of $\fc$ with phonons in drum 2 at a frequency of $\fmtwo$~\cite{Palomaki2013}.  If the TMS and BS interactions were applied separately, the former would energize drum 1 and the latter would cool drum 2~\cite{Kotler:2020SI}. However, because we apply the TMS and BS pulses simultaneously, energy flows to both drums. Thus, the cavity mediates the interaction between the drums, in a manner that is similar to other theoretical proposals~\cite{Mancini2002,Pinard_2005,Hartmann2008,Hofer2011,Tan2013,Wang2013,Vitali2015,Buchmann2015,Schnabel2015}.  As a result, strong correlations form between the quadratures of motion.

\N
Measurement is performed by amplifying a reflected microwave readout pulse after it has interacted with the drums. Typical microwave measurement efficiencies are $\sim 0.01$ even when using the best commercially available low-noise amplifiers. Here we achieve higher effective efficiency by using the TMS interactions native to the device as a preamplifier, following techniques that were developed for single drum readout~\cite{Palomaki710,Lecocq2015,Reed2017,Delaney2019}. We extend these methods, using frequency multiplexing, and improve our measurement efficiencies by more than an order of magnitude: $\eta_1=\etaone$, and $\eta_2=\etatwo$. Ultimately, the Heisenberg uncertainty principle prevents these efficiencies from exceeding $\onehalf$ since they quantify a concurrent measurement of both quadratures of motion~\cite{Caves:1980}. 

%
\begin{figure} 
	\centering
	\includegraphics[scale=1]{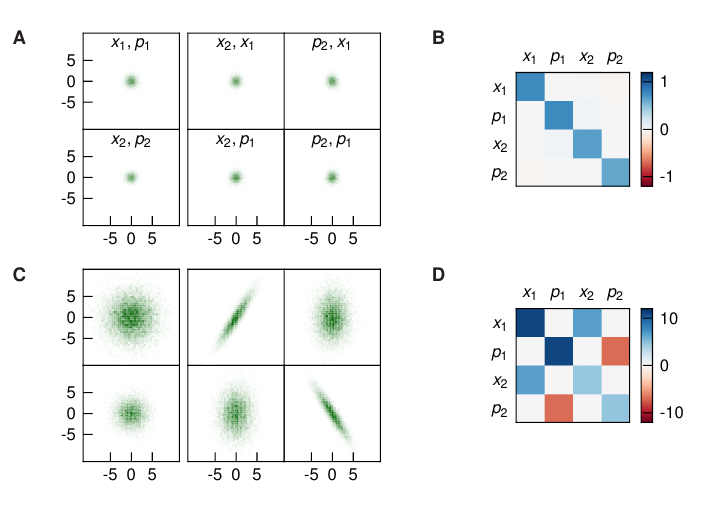}
	\caption{
		{\bf Tomography of two mechanical oscillators.}
		\textbf{(A)}  Histograms of a sideband-cooled state of two drums. Each experiment records the system variables $\vec{s}=\left(x_1,p_1,x_2,p_2\right)$ concurrently. Variables are scaled to dimensionless units according to the canonical commutation relation $[x_j,p_j]=i$ for $j=1,2$, so the vacuum state has variance $\onehalf$. Panel with legend $s_j,s_k$ corresponds to a correlation of $s_j$ along the $x$-axis and $s_k$ along the $y$-axis, quantified using a normalized 2d-histograms of $10,000$ experiment repetitions. The drums' individual variances are $\Vone=\nonecold$ and $\Vtwo=\ntwocold$.
		\textbf{(B)} Covariance matrix of the data in A.
		\textbf{(C)} Histograms of an entangled state of two drums. After sideband-cooling, a $\SI{16.8}{\micro\second}$ entangling pulse generates $x_1,x_2$ correlation and $p_1,p_2$ anti-correlation. Since the entanglement pulse pumps energy into the two-drum system, each drum's individual variance grows from their ground state cooled value to $\Vone=\nonehot$ and $\Vtwo=\ntwohot$, respectively.
		\textbf{(D)} Covariance matrix of the data in C. Correlations and anti-correlations are apparent in the off-diagonal elements. 
	}
\end{figure}

\N
Figure 2 shows tomography of the two-drum system, as characterized by its covariance matrix, for different protocols. First, we prepare a fiducial cold state by applying a pulse sequence of ground state cooling followed by readout, rendering a single concurrent measurement of $x_1,p_1,x_2,p_2$. Fig.~2A shows the experimental distribution of the measured variables for $10,000$ repetitions of the experiment. The two-dimensional histograms show no correlation between any of the measured variables. This is reaffirmed by the fact that the covariance matrix of the state is diagonal to a good approximation (Fig.~2C). The magnitude of the diagonal elements correspond to nearly ground-state variances of $\Vone=\nonecold$ and $\Vtwo=\ntwocold$ for drum 1 and 2 respectively, where  $V_j=\onehalf(\langle (x_j-\langle x_j\rangle)^2+ (p_j-\langle p_j\rangle)^2\rangle)$ for $j=1,2$. We now turn to a pulse sequence of ground state cooling,  entanglement and readout. Fig.~2B exhibits all the expected features of a highly correlated state. First, the $x,p$ histogram for each drum is consistent with a Gaussian distribution of large variance. Drum 1, which undergoes a TMS interaction, has a bigger variance ($\Vone=\nonehot$) than that of drum 2 ($\Vtwo=\ntwohot$), which undergoes a BS interaction. Second, a clear signature of drum-drum interaction is demonstrated by the correlation of $x_1,x_2$ and the anti-correlation of $p_1,p_2$.  The covariance matrix of the measured variables in Fig.~2D displays a dominant diagonal and four off-diagonal elements $\cC_{1,3}\approx -\cC_{2,4}$ and $\cC_{3,1}\approx -\cC_{4,2}$. Indeed, these clear correlations are directly observable in the  measured variables. Delineating them from classical correlations requires an application of the Simon-Duan criteria.
 
 \begin{figure} 
 	\centering
 	\includegraphics[scale=1]{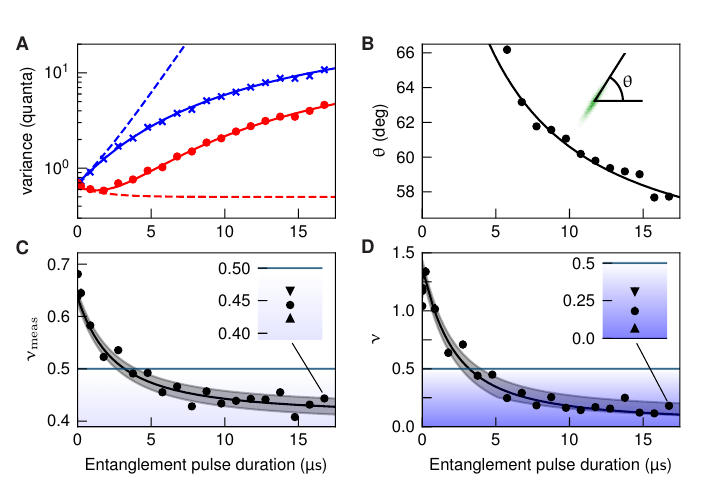}
 	\caption{
 		{\bf Entanglement of two drums versus pulse duration.} 
 		\textbf{(A)} Individual drum measured variances. Entangling is comprised of an energizing pulse for drum 1 and a cooling pulse for drum 2, applied simultaneously. The dashed lines show a theoretical prediction of the drums' individual variances if energizing and cooling where employed separately. Blue and red marks are the measured variances $V_1$ and $V_2$ for drums 1 and 2 respectively. Solid lines in all panels show theory with parameters obtained by fitting to an independent data set and without further adjustment~\cite{Kotler:2020SI}. 
 		\textbf{(B)} Angle of the $x_1,x_2$ correlation that determines the squeezed and anti-squeezed joint quadratures of the bipartite system.
 		\textbf{(C)} Entanglement in the measured variables, after loss, quantified by $\numeas$. Points below $\onehalf$ indicate entanglement of the two drums. Statistical error bars, quantified by 1-sigma bias-corrected bootstrapping confidence intervals, are smaller than the markers for most points. Shaded gray area corresponds to a 1-sigma uncertainty region in the location of the black theoretical prediction curve caused by measurement efficiency uncertainties. Inset shows the systematic uncertainty ($\pm$1-sigma) of the last measured point in the main graph (circle), indicated by the upper and lower points (triangles). The last point attains $\numeas=\reportednumeas$. 
 		\textbf{(D)} Entanglement in the mechanical variables, prior to loss, quantified by $\nu$. Uncertainties and inset plot are similar to C. The last point attains $\numech=\reportednumech$.
 	}
 \end{figure}
 
\N
Evolution of the two-drum state is shown for different entangling pulse durations  (Fig.~3), all of which are kept significantly shorter than $\SI{100}{\micro\second}$ to avoid the thermal  decoherence of both drums~\cite{Kotler:2020SI}. First, we focus on the individual variances of each drum (Fig.~3A). At short times $<\SI{1}{\micro\second}$, the drums show no evidence of interaction. Drum 2 cools while drum 1 becomes energized, the same behavior that would have been expected if the drums did not interact with one another. The dashed lines in Fig.~3A extrapolate this non-interacting behavior. Entangling pulse durations of $>\SI{1}{\micro\second}$ result in both drums deviating from this independent evolution; their variances now grow together in time with similar rates. Second, recall that in Fig.~2C, the histogram plot exhibited a correlation between  $x_1$ and $x_2$. Theory predicts that the angle that the elliptical $x_1,x_2$ distribution's major axis makes with the horizontal will evolve as shown by the solid theory line in Fig.~3B. Third, we focus on the measured variable $\numeas$, shown in Fig.~3C. Because our cooling of the drums is imperfect, their initial state contains some residual thermal motion in addition to the quantum fluctuations. The entangling operation must overcome this classical noise before the drums can be truly entangled. This is why, for pulse durations shorter than $\sim \SI{4}{\micro\second}$, $\numeas>\onehalf$ which strongly suggests that the drums are only classically correlated. For longer pulse durations, true quantum behavior, inconsistent with classical correlation, is observed and $\numeas$ crosses below $\onehalf$, indicating entanglement. At $\SI{16.8}{\micro\second}$ interaction time we observe $\numeas=\reportednumeas$, where ``sys" indicates systematic uncertainty dominated by uncertainty in the measurement efficiencies, and ``stat" indicates statistical uncertainty estimated using bootstrapping~\cite{Kotler:2020SI}. This is a direct measurement of entanglement for a bipartite system of macroscopic objects. It quantifies the amount of entanglement left in the system after noise processes have intervened during the measurement, and is therefore important for future quantum information applications. The amount of entanglement prior to the effect of noise can be estimated as well. To that end, we use a semi-definite program to find the closest (in $l_2$ distance) physically-realizable covariance matrix that, after an interaction with noise, is consistent with the data. From that we estimate the entanglement criterion $\numech$, which is shown in Fig.~3D. Indeed, we see more entanglement for the same interaction time, with $\numech=\reportednumech$. Such a level of entanglement might be useful for the exploration of mesoscopic Einstein-Podolsky-Rosen non-locality~\cite{Reid2019} and fundamental tests of quantum mechanics~\cite{Schnabel2015}.

\N
Our results demonstrate pulsed, time-domain control of three important building blocks for CV quantum information processing and quantum communication: state initialization, entanglement and measurement. Pulsed control played a key role. It allowed optimization of each piece separately and improved our measurement efficiency by more than an order of magnitude compared to traditional steady-state operation. As a result, we generated a highly entangled state of two macroscopic mechanical oscillators, surpassing the entanglement threshold by $\reportednumechdB~\textrm{dB}$. Most excitingly, we observe entanglement directly in the measured variables. This is relevant to future applications that require decisions based on measurement outcomes. We therefore expect the methods described here to serve as a stepping stone for teleportation and entanglement swapping of states of massive objects. This would enable novel hybrid quantum networks, where mechanics entangles with microwave fields~\cite{Palomaki710}, with spin systems~\cite{Thomas:2020} or is used as an intermediary to entangle radiation~\cite{Barzanjeh2019,Chen2020}.

\nocite{BowenBook,Simon2000,Duan2000,HorodeckiRevModPhys2009,Adesso:2014,Teufel2011,Teufel2011coolng,Palomaki2013,Palomaki710,Werner:2001gm,scully_zubairy_1997,Palomaki2013,Palomaki710,Delaney2019,Cicak:2010,Aspelmeyer:2014ce,Reed2017,Efron:1994}
\nocite{Buckland:1984}
\renewcommand{\refname}{REFERENCES AND NOTES}

\section*{ACKNOWLEDGMENTS} 
We thank Boaz Katz and Danny Ben-Zvi for feedback and insight on data taking and analysis. We thank Bradley Hauer and Adam Sirois for their careful reading of the manuscript. We thank Konrad Lehnert and Robert Delaney for useful discussions on measurement efficiency. We thank Noa Kotler for consulting on data and concept visualization. \textbf{Funding:} At the time this work was performed, S.K., E.S. F.L., A.K., and S.Geller were supported as Associates in the Professional Research Experience Program (PREP) operated jointly by NIST and the University of Colorado Boulder under Award No. 70NANB18H006 from the U.S. Department of Commerce. \textbf{Author contribution:} J.D.T. and S.K. designed the experiment. S.K. and G.A.P. fabricated the device. F.L. and K.C. supervised device fabrication. S.K. performed the measurements. F.L., R.W.S. and J.A. advised on measurement techniques. S.K., E.S., A.K. and S.Geller developed the theory and wrote, analyzed and tested the data analysis code.  S.K. and E.S. analyzed the results. S.Glancy and M.K. supervised theory work and data analysis. S.K., G.A.P., F.L., K.C., R.W.S., J.A. and J.D.T. developed the experimental infrastructure necessary to conduct the experiment. J.D.T. supervised the work. All authors provided experimental suggestions, discussed the results, and contributed to the writing of the manuscript. \textbf{Competing interests:} The authors declare no competing interests. S.K. is also affiliated with Qedma Quantum Computing Ltd. G.A.P. is also affiliated with PsiQuantum. \textbf{Data and materials availability:} All data are available in the manuscript or the supplementary material. This is a contribution of the National Institute of Standards and Technology, not subject to U.S. copyright.

\newpage
\section*{SUPPLEMENTARY MATERIALS}
Supplementary Text \newline
Figs. S1 to S4\newline
Tables S1 to S3\newline
References \textit{(34-42)}

\tableofcontents

\def\theequation{S\arabic{equation}}
\def\thefigure{S\arabic{figure}}
\def\thetable{S\arabic{table}}
\makeatother

\section{Continuous variable entanglement criterion}
For a continuous variable system such as ours, the state can be expressed in terms of the moments of the quadratures of motion observables. We use the dimensionless $X_j=(b_j^\dagger+b_j)/\sqrt{2}$ and $P_j=i(b_j^\dagger-b_j)/\sqrt{2}$ such that $[X_j,P_j]=i$, for modes $j=1,2$. Knowledge of all the moments formed by these variables constitutes a complete description of the system state. In this paper, however, we focus only on first and second moments since those give sufficient information to prove entanglement, as detailed below. 
We form the $4\times 4$ covariance matrix $\pazocal{C}$ of $\vec{S}=\left(X_1,P_1,X_2,P_2\right)$, such that element $j,k$ is $\pazocal{C}_{jk}=\tfrac{1}{2}\langle (S_j-\langle S_j\rangle)(S_k-\langle S_k\rangle)+(S_k-\langle S_k\rangle)(S_j-\langle S_j\rangle)\rangle$ where $\langle\ldots\rangle$ denotes the expectation value. Our entanglement criterion (Simon-Duan) can be calculated by subdividing $\pazocal{C}=\left(\begin{smallmatrix}
	A&C\\C^t&B
\end{smallmatrix}\right)$ where $A,B,C$ are $2\times 2$ sub-matrices, $C^t$ the latter's transpose, and using the formula, 
\begin{equation}\label{eq:numech}
	\numech = \sqrt{\tfrac{1}{2} \left(\Delta - \sqrt{\Delta^2-4\det\pazocal{C}}\right)}, 
\end{equation}
where $\Delta=\det(A)+\det(B)-2\det(C)$ (see~\mycite{BowenBook}). The formulation of the Simon-Duan criterion~\mycite{Simon2000,Duan2000} we use in our paper is:
\begin{equation}\label{eq:simon-duan}
	\numech <\tfrac{1}{2}\Rightarrow \textrm{entanglement}.
\end{equation}
Therefore we use Eq.~\eqref{eq:simon-duan} in the paper as a sufficient condition to show entanglement, based on first and second moments only. 

Although the theory behind Eq.~\eqref{eq:numech} and~\eqref{eq:simon-duan} is well known, it is worth recalling that it is based on the Peres-Horodecki positive partial transpose criterion for entanglement~\mycite{HorodeckiRevModPhys2009}. This criterion is stated in terms of  a density matrix $\rho$ of a bipartite system and its partial transpose $\rho^{\mathrm{pt}}$. If  $\rho$ describes a separable state then $\rho^{\mathrm{pt}}$ is a valid density matrix. Therefore, if $\rho^{\mathrm{pt}}$ is not a valid density matrix (no longer positive), then $\rho$ describes an entangled state. In the case of oscillators, $\numech$ is the minimal symplectic eigenvalue of the covariance matrix $\pazocal{C}^\textrm{pt}$ of the partial transposition of the oscillator's density matrix $\rho^{\mathrm{pt}}$ .  When a density matrix is partially transposed, the $P_2$ quadrature of the transposed mode is reversed, so $\pazocal{C}^\textrm{pt}=\Lambda\pazocal{C}\Lambda$, for $\Lambda=\left(\begin{smallmatrix}
	1&0&0&0\\
	0&1&0&0\\
	0&0&1&0\\
	0&0&0&-1
\end{smallmatrix}\right)$. It can be shown~\mycite{Adesso:2014} that if $\rho^{\mathrm{pt}}$ is a valid density matrix, then its minimal symplectic eigenvalue $\numech\ge \tfrac{1}{2}$. Therefore, if $\rho$ is separable then $\numech\ge\tfrac{1}{2}$. This implies Eq.~\eqref{eq:simon-duan}.


\subsection{A short discussion on Gaussian states}
It is important to stress that our entanglement criteria does not rely on the assumption that the system state is Gaussian. Nevertheless, our intuition and experience are based on Gaussian states and therefore warrant a short discussion.  

Based on both theory and experiments~\mycite{Teufel2011,Teufel2011coolng,Palomaki2013,Palomaki710}, the following two assumptions  are reasonable: (1) The initial state is Gaussian (a low-temperature thermal state). (2) State evolution is governed by bi-linear interactions (beam-splitters and two-mode squeezers). Therefore, the experiment evolution prior to measurement is confined to Gaussian states. As seen in Fig.~2 of the paper, these assumptions are consistent with the experimental data. The theory plots in Fig.~3 assume that the initial state of the mechanical elements is Gaussian and are therefore confined to Gaussian states (see Section~\ref{sec:theory}). That the curves are in reasonable agreement with the data is further evidence of Gaussian-confined evolution of the experiment.

Knowing that the state of the system is Gaussian allows stronger claims than the ones in this paper. First, a Gaussian state is completely determined by its first and second moments. Second, Eq.~\eqref{eq:simon-duan} becomes a necessary and sufficient condition for $1\times 1$-Gaussian states\footnote{Gaussian states that describe a bipartite system, with each sub-system having a single canonical degree of freedom. See~\mycite{Werner:2001gm}.} as shown by Simon and Duan~\mycite{Simon2000,Duan2000}. Therefore, if $\numech\ge\tfrac{1}{2}$ then the state is separable. Since proving that a state is separable was not the focus of this paper, we did not need to assume or quantify the Gaussianity of our states.

\section{Theory of entanglement versus time}~\label{sec:theory}
This section describes the theory relating the growth of entanglement to the entangling pulse duration as plotted in Fig.~3 of the paper. 
\subsection{Assumptions and derivation}
Our system is composed of three harmonic oscillators: two mechanical drums and one microwave cavity. The free evolution is determined by the Hamiltonian:
\begin{equation}
	H_0 = \hbar\omega_{\textrm{c}} a^\dagger a + \hbar\Omega_{\textrm{m,1}} b_1^\dagger b_1 +\hbar\Omega_{\textrm{m,2}} b_2^\dagger b_2,
\end{equation}
where $\omega_{\textrm{c}}=2\pi f_{\textrm{c}}$ is the radial frequency of the microwave cavity and $\Omega_{\textrm{m,j}}=2\pi f_{\textrm{m,j}}$ for $j=1,2$ are the radial frequencies of mechanical oscillators 1 and 2.

We move to a rotating frame with respect to $H_0$. In this frame, the system evolves according to the interaction Hamiltonian:
\begin{equation}\label{eq:Hint}
	H_\textrm{int} = \hbar g_b (a^\dagger b_1^\dagger+a b_1) + \hbar g_r(a^\dagger b_2 +a b_2^\dagger),
\end{equation}
which includes two electromechanical\footnote{The usage of the term ``electromechanical'' is meant to articulate the fact that we use optomechanical-type interactions in the microwave frequency domain.} interactions: (1) Two-mode squeezing (TMS) between the cavity and drum 1 with an interaction strength  $g_b$ and (2)  Beam splitter (BS) between the cavity and drum 2 with an interaction strength $g_r$. Derivations of electromechanical interactions have been done multiple times in the literature (for example, see~\mycite{Teufel2011coolng}) . The magnitudes of $g_b$ and $g_r$ are controlled using microwave pump pulses: $g_j=g_{0,j}\alpha_j^\textrm{coh}$ for $j\in\{b,r\}$ where the $\lvert\alpha_j^\textrm{coh}\rvert$'s are the amplitudes of the pump coherent states and $g_{0,j}$'s denote the bare electromechanical couplings. In a microwave pulse, $\alpha_j^\textrm{coh}=\alpha_j^\textrm{coh}(t)$ has a time-dependent envelope built from a rise time, a plateau and a fall time. For the purpose of the derivation here, we assume $\alpha_j^\textrm{coh}$, and therefore $g_j$, are constant and real. Generalization to the case where $\alpha_j^\textrm{coh}$ is a complex number is straightforward. 

We use the Heisenberg-Langevin formalism~\mycite{scully_zubairy_1997} to form the equations of motion, assuming that the microwave cavity has a radial damping rate of $\kappa$ and that the mechanical oscillators are dissipation-less. The latter holds since we work in a regime where  all the rates in the problem are much faster than the mechanical decoherence rates. This is consistent with the experimental parameters extracted in subsection~\ref{subsec:theory_params} and subsection~\ref{subsec:decoherence}: the entanglement rate is approximately a factor of 10 faster than the mechanical decoherence rate. The equations of motion therefore take the following simple form:
\begin{align}
	\dot{a} &= -\frac{\kappa}{2} a +\sqrt{\kappa}a_\textrm{in}-ig_bb_1^\dagger -ig_rb_2,\\
	\dot{b_1^\dagger} &=ig_ba,\label{eq:b1daggerdot}\\
	\dot{b_2} &=-ig_r a,\label{eq:b2dot}
\end{align} 
where $a_\textrm{in}$ is the cavity input noise operator satisfying $\langle a_\textrm{in}^\dagger(t_1) a_\textrm{in}(t_2)\rangle=n_{\textrm{in}}\delta(t_1-t_2)$.

To solve the equations of motion, first we assume that the cavity is effectively in steady state, i.e that $\dot{a}=0$. This is a good approximation as shown below. Therefore, we can express,
\begin{equation}\label{eq:aequals}
	a = \frac{2}{\kappa} \left(\sqrt{\kappa}a_\textrm{in}-ig_bb_1^\dagger -ig_rb_2\right), 
\end{equation}
and substitute the right hand side of  Eq.~\eqref{eq:aequals} for $a$ in the equations for $b_1^\dagger$ and $b_2$ (Eq.~\eqref{eq:b1daggerdot} and Eq.~\eqref{eq:b2dot}).  Solving for $b_1^\dagger$ and $b_2$ and expressing the solution in terms of the quadratures of motion,  $X_j=(b_j+b_j^\dagger)/\sqrt{2}$, $P_j=i(b_j^\dagger-b_j)/\sqrt{2}$ for $j\in\left\{1,2\right\}$ and the cavity input noise quadratures $I_{\textrm{in}} = (a_{\textrm{in}}^\dagger+a_{\textrm{in}})/\sqrt{2}$ and $Q_{\textrm{in}} = i(a_{\textrm{in}}^\dagger-a_{\textrm{in}})/\sqrt{2}$, we get:

\begin{align}
	X_1(t)& =\frac{\Gamma_r-\Gamma_bf(t)}{\Gamma_r-\Gamma_b}X_1(0)+\frac{\sqrt{\Gamma_r\Gamma_b}}{\Gamma_r-\Gamma_b}\left(1-f(t)\right)X_2(0)\nonumber \\
	&-\sqrt{\Gamma_b}\int_{0}^td\tau f(\tau) Q_{\textrm{in}}(t-\tau),\\
	X_2(t)& = \frac{\Gamma_rf(t)-\Gamma_b}{\Gamma_r-\Gamma_b}X_2(0)-\frac{\sqrt{\Gamma_r\Gamma_b}}{\Gamma_r-\Gamma_b}\left(1-f(t)\right)X_1(0)\nonumber \\
	&+\sqrt{\Gamma_r}\int_{0}^td\tau f(\tau) Q_{\textrm{in}}(t-\tau),\\
	P_1(t)& = \frac{\Gamma_r-\Gamma_bf(t)}{\Gamma_r-\Gamma_b}P_1(0)-\frac{\sqrt{\Gamma_r\Gamma_b}}{\Gamma_r-\Gamma_b}\left(1-f(t)\right)P_2(0) \nonumber \\
	&-\sqrt{\Gamma_b}\int_{0}^td\tau f(\tau) I_{\textrm{in}}(t-\tau),\\
	P_2(t)& = \frac{\Gamma_rf(t)-\Gamma_b}{\Gamma_r-\Gamma_b}P_2(0)+\frac{\sqrt{\Gamma_r\Gamma_b}}{\Gamma_r-\Gamma_b}\left(1-f(t)\right)P_1(0)\nonumber \\
	&-\sqrt{\Gamma_r}\int_{0}^td\tau f(\tau) I_{\textrm{in}}(t-\tau),
\end{align}

where
\begin{align}\label{eq:gammadef}
	f(t)&=\exp\left(\frac{\Theta(t)}{2}\right),\nonumber\\
	\Theta(t)&=(\Gamma_b-\Gamma_r)t,\nonumber\\
	\Gamma_b&=\frac{4g_b^2}{\kappa},\nonumber\\
	\Gamma_r&=\frac{4g_r^2}{\kappa}.
\end{align}
Two comments about these solutions are in order. First, recall the assumption that $g_r$ and $g_b$ are time-independent. This leads to $\Theta$ depending linearly on $t$. If, however,  $g_r$ and $g_b$ are time-dependent, then the solution can be approximated by setting $\Theta(t) = \int_0^t dt' (\Gamma_b(t')-\Gamma_r(t'))$.  Second, note that the assumption leading to Eq.~\eqref{eq:aequals} is self-consistent since experimentally, $\kappa/2\pi\sim\SI{800}{\kilo\hertz}$ and $\Gamma_r/2\pi,\Gamma_b/2\pi<\SI{100}{\kilo\hertz}$ (see subsection~\ref{subsec:theory_params}). 

We now use the solution for the quadratures of motion to find the time dependent covariance matrix (second moments). We assume that the state of the drums at $t=0$ is a separable thermal state characterized by the variances $\langle X_1^2(0)\rangle=\langle P_1^2(0)\rangle=(n_b+1/2)$ and $\langle X_2^2(0)\rangle=\langle P_2^2(0)\rangle=(n_r+1/2)$ where $n_b$ and $n_r$ are the respective mean thermal occupancies. The resulting covariance matrix at time $t$ has the form:
\begin{equation}\label{eq:cov}
	\pazocal{C} = \begin{pmatrix}
		\alpha(t) & 0          & \gamma(t) & 0          \\
		0         & \alpha(t)  & 0         & -\gamma(t) \\
		\gamma(t) & 0          & \beta(t)  & 0          \\
		0         & -\gamma(t) & 0         & \beta (t)
	\end{pmatrix},
\end{equation}

where
\begin{align*}
	\alpha(t) = &(n_b+\tfrac{1}{2}) \left(\frac{\Gamma_r - \Gamma_b\f}{\Gamma_r-\Gamma_b}\right)^2
	+ (n_r+\tfrac{1}{2})\frac{\Gamma_r\Gamma_b}{(\Gamma_r-\Gamma_b)^2}(1-\f)^2\\
	&+(n_{\textrm{in}}+\tfrac{1}{2})\frac{\Gamma_b}{\Gamma_r-\Gamma_b}(1-\fsq),\\ 
	\beta(t) =& (n_r+\tfrac{1}{2}) \left(\frac{\Gamma_r\f-\Gamma_b}{\Gamma_r-\Gamma_b}\right)^2 
	+(n_b+\tfrac{1}{2}) \frac{\Gamma_r\Gamma_b}{(\Gamma_r-\Gamma_b)^2}(1-\f)^2 \\
	&+(n_{\textrm{in}}+\tfrac{1}{2}) \frac{\Gamma_r}{\Gamma_r-\Gamma_b}(1-\fsq),\\ 
	\gamma(t) = &\frac{\sqrt{\Gamma_r\Gamma_b}}{\Gamma_r-\Gamma_b}\bigg[\left((n_r+\tfrac{1}{2})\frac{\Gamma_r\f-\Gamma_b}{\Gamma_r-\Gamma_b}-(n_b+\tfrac{1}{2}) \frac{\Gamma_r-\Gamma_b\f}{\Gamma_r-\Gamma_b}\right)(1-\f)\\
	&-(n_{\textrm{in}}+\tfrac{1}{2})(1-\fsq)\bigg].
\end{align*}

For a covariance matrix of the form in Eq.~\eqref{eq:cov}, 
\begin{equation}\label{eq:nu}
	\nu = \sqrt{\tfrac{1}{2}\left(\alpha^2+\beta^2+2\gamma^2-(\alpha+\beta)\sqrt{(\alpha-\beta)^2+4\gamma^2}\right)}.
\end{equation}

\subsection{Theory for measured variables}
The inefficiency of our measurement apparatus can be modeled as an effective beam-splitter~\mycite{Palomaki2013,Palomaki710,Delaney2019}, where each quadrature variable describing the drums $S_i\in\left\{X_1,P_1,X_2,P_2\right\}$, becomes mixed with vacuum noise such that, 
\begin{equation}\label{eq:beamsplitter}
	s_{i}=\sqrt{\eta_i}S_i +\sqrt{1-\eta_i}\xi_i,
\end{equation}
where $\eta_i$ is the efficiency of the measurement of drum $i$, $\xi_i$ is a Gaussian random variable with zero mean and vacuum variance $\langle \xi_i^2\rangle=\tfrac{1}{2}$, and $\xi_i$ and $\xi_j$ are independently distributed for $i\neq j$.

We form the measured covariance matrix $\pazocal{C}_\textrm{meas}$ from the measured variables $s_i$, and calculate the measured minimal symplectic eigenvalue $\numeas$ by applying formula~\eqref{eq:numech} to $\pazocal{C}_\textrm{meas}$:
\begin{equation}\label{eq:numeasgen}
	\numeas = \sqrt{\tfrac{1}{2} \left(\Delta_\textrm{meas} - \sqrt{\Delta_\textrm{meas}^2-4\det\pazocal{C}_\textrm{meas}}\right)}, 
\end{equation}
where $\Delta_\textrm{meas}=\det(A_\textrm{meas})+\det(B_\textrm{meas})-2\det(C_\textrm{meas})$ and $\pazocal{C}_\textrm{meas}=\left(\begin{smallmatrix}
	A_\textrm{meas}&C_\textrm{meas}\\C^t_\textrm{meas}& B_\textrm{meas}
\end{smallmatrix}\right)$.

If $\pazocal{C}$ has the special form as in Eq.~\eqref{eq:cov} then the resulting measured covariance matrix has the same form:
\begin{equation}\label{eq:covmeas}
	\pazocal{C}_\textrm{meas} = \begin{pmatrix}
		\alpha_{\textrm{meas}}(t) & 0 & \gamma_{\textrm{meas}}(t) & 0          \\
		0 & \alpha_{\textrm{meas}}(t) & 0 & -\gamma_{\textrm{meas}}(t) \\
		\gamma_{\textrm{meas}}(t) & 0 & \beta_{\textrm{meas}}(t) & 0          \\
		0 & -\gamma_{\textrm{meas}}(t) & 0 & \beta_{\textrm{meas}}(t)
	\end{pmatrix},
\end{equation}

where
\begin{align}\label{eq:alphameas}
	\alpha_{\textrm{meas}}(t) = &\eta_1\alpha(t) +(1-\eta_1)\frac{1}{2}, \nonumber \\
	\beta_{\textrm{meas}}(t) =& \eta_2\beta(t)+(1-\eta_2)\frac{1}{2} ,\nonumber \\
	\gamma_{\textrm{meas}}(t) = &\sqrt{\eta_1\eta_2}\gamma(t).
\end{align}

These elements of the measured covariance matrix correspond to $\numeas$, the measured minimal symplectic eigenvalue:
\begin{equation}\label{eq:numeas}
	\numeas = \sqrt{\tfrac{1}{2}\left(\alpha^2_\textrm{meas}+\beta^2_\textrm{meas}+2\gamma^2_\textrm{meas}-(\alpha_\textrm{meas}+\beta_\textrm{meas})\sqrt{(\alpha_\textrm{meas}-\beta_\textrm{meas})^2+4\gamma^2_\textrm{meas}}\right)}.
\end{equation}

What is the relationship between $\numech$ and $\numeas$ ? In the special case of a symmetric state with symmetric loss where $\alpha=\beta$ and $\eta_1=\eta_2=\eta$,  $\numeas$ becomes a convex combination of $\tfrac{1}{2}$ and $\numech$. By plugging the relations in Eq.~\eqref{eq:alphameas} into Eq.~\eqref{eq:numeasgen}, we get:
\begin{equation}\label{eq:convex}
	\numeas=\eta \numech+(1-\eta)\tfrac{1}{2},
\end{equation}
Therefore, $\numech<\tfrac{1}{2}$ if and only if $\numeas<\tfrac{1}{2}$. It turns out that this statement holds in the more general case:

\newtheorem*{claim}{Claim}
\begin{claim}
	For a state described by a covariance matrix of the form in Eq.~\eqref{eq:cov} the state after loss is entangled if and only if the state before loss is entangled:
	$\numech<\tfrac{1}{2}$ if and only if $\numeas<\tfrac{1}{2}$. 
\end{claim}
\begin{proof}
	First, we translate the Simon-Duan entanglement criterion in Eq.~\eqref{eq:simon-duan} to a different form. Using the expression for $\numech$ in Eq.~\eqref{eq:nu}, Eq.~\eqref{eq:simon-duan} is algebraically equivalent to:
	\begin{equation}\label{eq:stepone}
		(\alpha\beta-\gamma^2)^2<\tfrac{1}{4}\left((\alpha+\beta)^2+2(\gamma^2-\alpha\beta)\right)-\tfrac{1}{16}.
	\end{equation}
	By setting $Z\equiv\alpha\beta-\gamma^2$ and solving for a quadratic equation in $Z$, it follows that~\eqref{eq:stepone} is equivalent to,
	\begin{equation}
		-\frac{1}{4}-\frac{\alpha+\beta}{2}<Z<-\frac{1}{4}+\frac{\alpha+\beta}{2},
	\end{equation}
	which in turn is equivalent to:
	\begin{equation}
		(\alpha-\tfrac{1}{2})(\beta-\tfrac{1}{2})<\gamma^2<(\alpha+\tfrac{1}{2})(\beta+\tfrac{1}{2}).
	\end{equation}
	The Cauchy-Schwarz inequality ensures that $\gamma^2 \le \alpha\beta$ for all covariance matrices, so only the left hand side is relevant for the entanglement test. We are left with the following criterion: A Gaussian state is entangled iff
	\begin{equation}
		\textrm{Ent}\equiv (\alpha-\tfrac{1}{2})(\beta-\tfrac{1}{2})-\gamma^2 < 0.
	\end{equation}
	With this new form of the Simon-Duan criterion, and using Eq.~\eqref{eq:alphameas}, we get:
	\begin{equation}
		\textrm{Ent}_\textrm{meas}=\eta_1\eta_2\textrm{Ent}.
	\end{equation}
	Therefore $\textrm{Ent}<0$ iff $\textrm{Ent}_\textrm{meas}<0$. 
\end{proof}

\subsection{Fitting of theory parameters}\label{subsec:theory_params}
We assume that $n_{\textrm{in}}=0$, i.e. that the cavity input noise operators, $I_\textrm{in}(t)$ and $Q_\textrm{in}(t)$, have vacuum occupancies: $\langle I(t_1)I(t_2)\rangle=\langle Q(t_1)Q(t_2)\rangle=\tfrac{1}{2}\delta(t_1-t_2)$. This is a reasonable assumption for our experiment that was run at a cryostat temperature of $\sim\SI{7}{\milli\kelvin}$ and used a microwave cavity with a  frequency of $\sim\SI{6}{\giga\hertz}$. The thermal occupancy in this case is negligible. In practice, both our experiment and previous work~\mycite{Teufel2011,Teufel2011coolng,Palomaki2013,Palomaki710} are consistent with $n_{\textrm{in}}\lesssim 0.1$. 

The rest of the theory parameters are $n_b$, $n_r$, $\Gamma_b$, $\Gamma_r$, $\eta_1$ and $\eta_2$. We use an independently measured data set to simultaneously fit (least-squares) for $\alpha_{\textrm{meas}}(t)$ and $\beta_{\textrm{meas}}(t)$ versus time $t$. From the fit we extract $n_b=\nb$, $n_r=\nr$, $\Gamma_b=\Gammab$, $\Gamma_r=\Gammar$, $\eta_1=\etaone$ and $\eta_2=\etatwo$, where parentheses denote 1-sigma confidence intervals. We stress two important points: (1) Only the thermal variances of drum 1 and drum 2 were used for the fit. We did not use the cross-correlation term $\gamma_{\textrm{meas}}$ nor the minimal symplectic eigenvalue for the fit. (2) The data in Fig.~3 was not used for the fit, and the theory plotted there is plotted without any further adjustment. 

\subsection{Theory plots in Fig 3}
\newcommand{\Var}{\mathrm{Var}}
The plot in Fig.~3A displays $\alpha_\textrm{meas}(t)$ and $\beta_\textrm{meas}(t)$, the drum variances versus entanglement pulse duration. The theoretical model assumes that each drum has equal quadrature variances, that is $\Var(x_1)=\Var(p_1)=\alpha_\textrm{meas}$ and $\Var(x_2)=\Var(p_2)=\beta_\textrm{meas}$.  However the estimates of each drum's quadrature variances, obtained from data, are not exactly equal.  Each data point in Fig. 3A shows the average of a drum's two estimated variances.

Figure 3B shows the correlation angle, i.e. the elliptical $x_1,x_2$ distribution's major axis makes with the horizontal:
\begin{equation}
	\theta= \arctan\left(\frac{\alpha_\textrm{meas}-\beta_\textrm{meas}-\sqrt{(\alpha_\textrm{meas}-\beta_\textrm{meas})^2+4\gamma_\textrm{meas}^2}}{2\gamma_\textrm{meas}}\right).
\end{equation}

The theory in Fig.~3C and Fig.~3D show the Simon-Duan criteria using Eq.~\eqref{eq:numeas} and~\eqref{eq:nu} respectively. We stress that the data points in Fig.~3C and Fig.~3D that plot $\numeas$ and $\numech$ were extracted from the experimentally obtained covariance matrix and \emph{without} assuming that it has the special form in Eq.~\eqref{eq:cov} and~\eqref{eq:covmeas}. Section~\ref{sec:dataanalysis} elaborates on how. 

\subsection{A short discussion of the entanglement protocol}
The entanglement protocol involves a simultaneous application of a TMS interaction between the cavity and drum 1 and a BS interaction between the cavity and drum 2. It is instructive to compare our simultaneous protocol with a sequential one: a TMS pulse followed by a BS pulse. 

To understand the appeal of the sequential protocol we start with the ideal case where neither the cavity nor the mechanical modes have any dissipation. The TMS would generate a two-mode squeezed state between drum 1 and the cavity. This would create strong correlations that grow exponentially in the TMS pulse duration $t_\textrm{TMS}$. At this point drum 1 and the cavity share an entangled state that is separable from drum 2. Then, a BS pulse can be used to swap the excitation in the cavity to drum 2, thereby transferring the quantum correlations to the drum 1-drum 2 system. Now the cavity state is separable from the entangled-drums state. In practice the cavity has some dissipation. The preferred protocol depends on this dissipation through the ratio $g/\kappa$. 

A sequential entangling protocol is preferable in the strong-coupling regime $g\gg \kappa$. In this case, the entangling rate becomes $\approx 2 g$ and therefore $t_\textrm{TMS}$ can be chosen to satisfy $1/(2g) < t_\textrm{TMS} < 1/\kappa$. This would ensure that strong correlations between drum 1 and the cavity can be generated before the effect of cavity dissipation becomes dominant. A similar argument holds for the BS pulse. In the strong coupling regime, most of the state of the cavity can be swapped into drum 2. Therefore, the amount of entanglement between the drums at the end of the protocol is bounded by the entanglement generated between drum 1 and the cavity mode at the intermediate state of the protocol.

Our device operates in the weak coupling regime $g< \kappa$. We intentionally designed for a large $\kappa$ by strongly coupling the electrical cavity to the output port. This facilitated fast readout that resolved the two frequency-multiplexed sidebands originating from the two drums with good fidelity (see Sec.~\ref{subsec:from_raw_data_to_units_of_quanta} and Fig.~\ref{fig:FFT}). As a result, the cavity dissipation rate was an order of magnitude larger than the maximal entangling rate we could practically attain (see table~\ref{tbl:parameters}). Pulsing the TMS and BS simultaneously mitigates this problem to a large extent. This is shown in the theory plot and data of Fig.~3D of the paper. To gain intuition, we consider the simple case of $n_b=n_r=n_\textrm{in}=0$ and $\Gamma_r=\Gamma_b=\Gamma\approx 4g^2/\kappa$ so that,
\begin{equation}
	\nu=\frac{1}{2}\sqrt{1-8r\left[(r^2+r+\frac{1}{2})\sqrt{r^2+2r+2} -(r^3+2r^2+2r+\frac{1}{2})\right]},
\end{equation}
for $r\equiv \Gamma t/2$. This function decreases monotonically from $\nu(r=0)=1/2$ to a limit value of $\lim_{r\to\infty}\nu=0$. Therefore, in the absence of mechanical dissipation, the drums may become maximally entangled, despite strong cavity dissipation. 

In our parameter regime applying the pulses in sequence would not have performed as well. First, the TMS pulse would render a weakly entangled state with a steady-state minimal symplectic eigenvalue of $\nu_\textrm{drum 1,cavity}\approx \frac{1}{2}\left(1-\frac{\Gamma}{\kappa}\right)$ where, we assumed $\Gamma_r=\Gamma_b=\Gamma$ for simplicity. Second, the BS pulse would then perform a swap between the cavity and drum 2 with a low efficiency of $\lesssim \Gamma/\kappa$ thereby diluting the entanglement correlations even further. We estimate that the combined effect of TMS followed by a BS pulse would have rendered a steady state value for $\nu\approx 0.495$ (or worse), very close to the $\tfrac{1}{2}$ threshold value for entanglement.

\section{Experimental details}
\subsection{Main challenges}
Our experiment relied on single drum addressing: the ability to apply a microwave control field (either for entanglement, state preparation or readout) mostly to a specific drum and measure the mechanical quadratures of each drum independently. As it turns out, achieving single drum addressing for a two-drum device while accommodating for other experimental constraints posed a significant challenge.

Since a single microwave cavity interfaced both drums, individual addressing relied on frequency multiplexing. We used a multi-tone pulse scheme where each tone could be associated with a specific drum.  This constrains the drum modes used in the experiment to have different mechanical frequencies, $\Omega_i$ for $i=1,2$, such that $\lvert \Omega_1 - \Omega_2 \rvert \gtrsim 3\kappa$, provided that their electromechanical coupling strengths are comparable.  Otherwise, a stronger pulse is required to address the weakly-coupled drum and could inadvertently actuate the strongly-coupled drum despite their frequency difference. Moreover, the presence of a strong pump  could saturate the device power handling capability and introduce nonlinear uncontrolled effects. 

Our original approach was to fabricate two drums with different diameters so that their fundamental frequencies are well resolved. These devices failed due to a thermal effect that influences the parallel plate separation $d$ of the drums~\mycite{Cicak:2010}. At room temperature, $d=\SI{200}{\nano\meter}$ nominally. After cooling the device, the plate separation ends at $d_\textrm{final}\sim35$ to $\SI{50}{\nano\meter}$.  While the process renders very similar $d_\textrm{final}$ values for identical drums, it can be very different for drums of different design in general, and drums of different diameters in particular. The ratio of the pulse powers required to operate the different drums is proportional to $(d_\textrm{final,1}/d_\textrm{final,2})^4$, where $d_\textrm{final,i}$ is the plate separation of drum $i$ at base temperature. Therefore, a small discrepancy in $d_\textrm{final}$ could render the device inoperable. 

We were able to solve this problem by fabricating drums that had an identical oblong top plate as detailed in subsection~\ref{subsec:device_detail}. Our implemented multi-tone frequency-domain and time domain multiplexing scheme that enabled single drum addressing is detailed in subsections~\ref{subsec:microwave_pulse_sequence} to~\ref{subsec:from_raw_data_to_units_of_quanta}.

\subsection{Device details}\label{subsec:device_detail}
Our two mechanical oscillators are made of lithographically-patterned thin-film aluminum that form drum-like membranes~\mycite{Teufel2011}, each with a mass of $\approx 70~\textrm{pg}$, suspended above a sapphire substrate, as shown in Fig.~1B. We use the $f_{\textrm{m},1}=10.9$ MHz mode of the left drum in Fig.~1B of the paper and the $f_{\textrm{m},2}=15.9$ MHz mode of the right drum. By design, the drums have no acoustic interaction with each other. Here we use electromechanics~\mycite{Aspelmeyer:2014ce} in order to mediate the interaction between the drums. Below each suspended drum, we fix an aluminum bottom plate to the substrate, so each drum is the top plate of a parallel-plate capacitor. A change in the distance between the drumhead and the bottom plate changes the capacitance of its corresponding capacitor. Both capacitors are shunted by a shared superconducting aluminum inductor, as shown in Fig.~1C  of the paper. The capacitors and inductor form a single microwave resonator, known as the `cavity', whose resonance frequency depends on the drums' motion, and is centered at $\fc=6.0806$ GHz. Thus, information about the drums can be encoded into the microwave field. We uniquely associate an acoustic frequency with a specific drum since their eigenmode frequencies are not degenerate and their bottom plates are split along the long and short axes of the drums (see Fig.~1B  of the paper). The latter changes the transduction strength between the drums' motions and the microwave cavity in a way that depends on the spatial overlap between the bottom plate and the acoustic mode shape, and therefore is drum-dependent. The cavity is inductively coupled to a coaxial line as shown in Fig.~1D of the paper, at a rate of $\approx 800$ kHz. 

\subsection{Electromechanics}
A microwave pulse, reflected off the cavity, imparts forces on the drums and encodes the amplitudes of their quadratures of motion into Doppler-shifted sidebands of the microwave pulse. The carrier frequency of the microwave pulse determines the nature of the interaction. A red-sideband (RSB) pulse is sent with a carrier frequency of $\fc-f_\textrm{m}$, where $f_\textrm{m}$ is the mechanical frequency of the drum of interest. It generates an effective beam-splitter (BS) interaction~\mycite{Palomaki2013}, $H_\textrm{RSB}= \hbar g_r(t) (ab^\dagger+ba^\dagger)$, where $g_r(t)$ is proportional to the pulse power at time $t$, $\hbar$ is the reduced Planck constant, $a$ and $a^\dagger$ are the annihilation and creation operators of the microwave cavity photons and  $b$ and $b^\dagger$ are the annihilation and creation operators of the drum's phonons, satisfying the canonical commutation relations of $[b,b^\dagger]=[a,a^\dagger]=1$. We use the RSB interaction to cool the mechanical mode nearly to the ground state~\mycite{Teufel2011coolng}. In contrast, a blue-sideband (BSB) pulse has a carrier frequency of $\fc+f_{m}$, resulting in an effective two-mode squeezing (TMS) interaction, $H_\textrm{BSB}=\hbar g_b(t) (a^\dagger b^\dagger + ba)$. These pulses entangle the motion of a mechanical oscillator with the itinerant reflected microwave field as well as allow for mechanical state readout~\mycite{Palomaki710}. As shown in the next subsection, RSB and BSB type pulses form all of the necessary ingredients for a complete experimental realization which includes ground-state cooling, entangling and state readout.

\subsection{Microwave pulse sequence}\label{subsec:microwave_pulse_sequence}
An experiment is composed of a sequence of pulses, each one defined by up to two carrier frequencies $f_1$ and $f_2$ and two amplitudes ($V_1$ for drum 1 and $V_2$ for drum 2)  and an envelope $A(t)$:
\begin{equation}
	A(t)\left(V_1\sin(2\pi f_1 (t-t_s) )+V_2\sin(2\pi f_2(t-t_s) )\right),
\end{equation}
where $t_s$ is the pulse start time. We used an envelope function of the form:
\begin{equation}
	A(t) = \left(1-\exp\left(-\frac{(t-t_s)^2}{2\tau_w^2}\right)\right) \left(1-\exp\left(-\frac{(t_e-t)^2}{2\tau_w^2}\right)\right),
\end{equation}
where $t_e$ is the pulse end time, and $\tau_w=\SI{1}{\micro\second}$ is the window rise/fall time. Since the pulses are stacked one after the other to form a single experiment, it is convenient to describe them in terms of pulse duration $t_e-t_s$. 

\begin{table}[!h]
	\caption{Sequence of pulses forming a single experiment. Microwave cavity resonant frequency is $\fc$. Mechanical frequencies are $f_{\textrm{m},1}$ and $f_{\textrm{m},2}$. See table~\ref{tbl:parameters} for measured values. }\label{tbl:pulses}
	\centering
	\begin{tabular}{llll}
		Name&  Duration&	\multicolumn{1}{c}{$f_1$}& \multicolumn{1}{c}{$f_2$}\\
		\hline
		Cooling&$\SI{100}{\micro\second}$& $\fc-f_{\textrm{m},1}+\SI{100}{\kilo\hertz}$& $\fc-f_{\textrm{m},2}-\SI{100}{\kilo\hertz}$\\
		Entangling&$0-\SI{20}{\micro\second}$& $\fc+f_{\textrm{m},1}$& $\fc-f_{\textrm{m},2}$\\
		Readout&$\SI{148}{\micro\second}$& $\fc+f_{\textrm{m},1}+\SI{200}{\kilo\hertz}$& $\fc+f_{\textrm{m},2}-\SI{200}{\kilo\hertz}$\\
		Reset&$\SI{640}{\micro\second}$& $\fc-f_{\textrm{m},1}+\SI{100}{\kilo\hertz}$& $\fc-f_{\textrm{m},2}-\SI{100}{\kilo\hertz}$\\
		Pilot&$\SI{1000}{\micro\second}$& $\fc+\SI{50}{\kilo\hertz}$&  \multicolumn{1}{c}{-}\\
		\hline
	\end{tabular}
\end{table}

Table~\ref{tbl:pulses} specifies the parameters that define the pulse sequence.  Frequency multiplexing is evident for the cooling, readout and reset pulses, since the electromechanical sidebands of the two drums do not overlap spectrally, by choosing two red sideband pump carrier frequencies that are $\SI{200}{\kilo\hertz}$ apart for cooling/reset and two blue sideband pump carrier frequencies that are $\SI{400}{\kilo\hertz}$ apart for readout. The latter implements phase-insensitive amplification for each of the drums. Figure 1E in the paper describes the first three steps of the sequence: cooling, entangling and readout. In addition, there are two auxiliary pulses: reset and pilot. The reset pulse cools the drums when they are highly energetic due to the entangling and BSB readout pulses. The pilot pulse has a carrier frequency that is almost resonant with the microwave cavity. By fitting it to a sine wave we extract the local oscillator phase for each individual experiment. This enables realigning different iterations of the experiment to  the same phase reference. 

The total experiment time is less than $\SI{1.9}{\milli\second}$. We repeat the experiment with a $\SI{10}{\milli\second}$ duty cycle. This avoids thermal and transient effects and ensures reproducible results. 
\setcounter{figure}{0}
\subsection{Tone generation and measurement chain}\label{subsec:tone_generation_and_measurement_chain}
\begin{figure}[hbt!]
	\centering
	\includegraphics{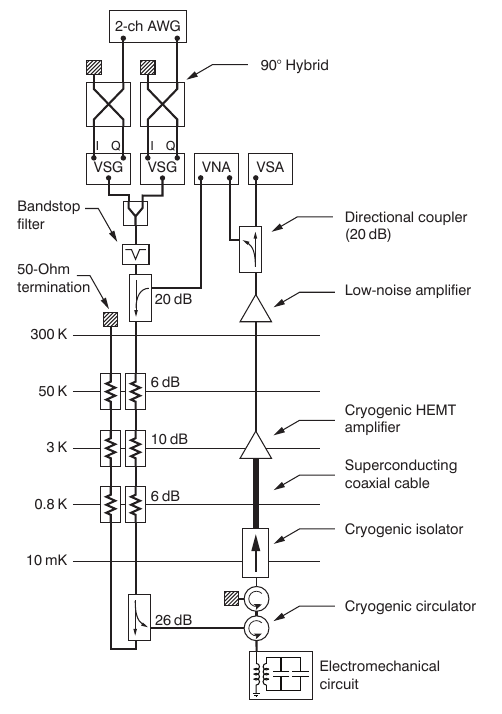}
	\caption{Measurement chain schematic. AWG: arbitrary waveform generator. VSG: vector signal generator. VNA: vector network analyzer. VSA: vector signal analyzer. A microwave pulse is sent through a room-temperature band-stop filter followed by more than $48~\textrm{dB}$ of cold attenuation to a circulator. The latter routes incoming pulses to the electromechanical circuit. A resulting reflected pulse that carries information about the mechanical motion of the drums is routed to an additional circulator, followed by an isolator. A High-Mobility Electron Transistor amplifier (HEMT) amplifies the signal. The signal is  amplified yet again at room temperature.  A VSA mixes the amplified microwave pulse to baseband and digitizes the result. Device diagnostics is done using a VNA.}\label{fig:measurement_chain}
\end{figure}
We performed our experiment in a dilution refrigerator, with a typical base temperature of $\sim\SI{7}{\milli\kelvin}$. Figure~\ref{fig:measurement_chain} shows detailed schematics of the tone generation and measurement chain. In broad strokes: a microwave pulse is generated at room temperature, travels into the dilution refrigerator, interacts with the electromechanical circuit at $\sim\SI{7}{\milli\kelvin}$, reflects back carrying information on the mechanical motion, and is finally demodulated and digitized at room temperature . 

Pulse generation was done by combining the output of two vector signal generators. We used the single-sideband-modulation technique for each generator so that with two generators, two sidebands could be combined simultaneously. Single sideband was implemented with an arbitrary waveform generator that outputs an intermediate frequency (IF) signal in the frequency range of $80-120~\SI{}{\mega\hertz}$ into a $90^\circ$-hybrid that feeds the $I$-$Q$ input of the generator. Both generators have their local oscillator (LO) set to $\fc-\SI{100}{\mega\hertz}$.  Therefore, the resulting microwave pulse has a bandwidth of $\sim\SI{40}{\mega\hertz}$ around $\fc$. 

Signal demodulation and digitization was done using a vector signal analyzer (VSA), where the demodulation was set to a $\SI{1}{\mega\hertz}$ bandwidth around $\fc$.

\subsection{From raw data to units of quanta}\label{subsec:from_raw_data_to_units_of_quanta}
We use blue-sideband readout techniques that were developed previously for single-drum devices~\mycite{Palomaki710,Reed2017}. Briefly, for each drum a blue-sideband microwave pump is pulsed during a time window of $t_s\le t\le t_e$. It produces a reflected microwave pulse with an exponentially increasing envelope, whose in-phase (I) and out-of-phase (Q) components are linear functions of the quadratures of motion of the drum at $t=t_s$:
\begin{align}\label{eq:IQofXP}
	I(t)  &= \sqrt{G} \sqrt{\Gamma_b} e^{\Gamma_b t/2}\left(\cos(\omega_\textrm{mod}t) X(t_s)-\sin(\omega_{\textrm{mod}}t)P(t_s)\right)+\xi_I(t),\nonumber\\
	Q(t) &= \sqrt{G} \sqrt{\Gamma_b} e^{\Gamma_b t/2}\left(\sin(\omega_\textrm{mod}t) X(t_s)+\cos(\omega_{\textrm{mod}}t)P(t_s)\right)+\xi_Q(t),
\end{align}
where $\Gamma_b$ is the blue sideband readout rate\footnote{Notice that we are abusing notation. The rate $\Gamma_b$ used here is the readout rate. It is independent and different from the rate used for entangling in Eq.~\eqref{eq:gammadef}. Unfortunately, there is only a finite number of Greek letters and too many pulse parameters.}, $\omega_{\textrm{mod}} = 2\pi f_\textrm{mod}$ and $f_\textrm{mod}=(f_\textrm{p}-\fc-f_\textrm{m})$ is the difference between the blue-sideband pump frequency $f_\textrm{p}$ and the sum of the cavity frequency $\fc$ and the mechanical frequency $f_\textrm{m}$, $G$ is the gain transduction factor that converts mechanical quanta units to voltage squared, and $\xi_I(t)$, $\xi_Q(t)$ denote readout noise. The VSA extracts the in-phase (I) and out-of-phase (Q) components of the reflected microwave pulse during the readout time window. The quadratures of motion $X(t_s)$ and $P(t_s)$ (plus relevant noise) are extracted using a straightforward linear filter applied to $I(t)$ and $Q(t)$. Applying the filter, as is evident from Eq.~\eqref{eq:IQofXP}, requires characterizing $\Gamma_b$, $\omega_{\textrm{mod}}$ and $G$.

While the blue-sideband readout technique was developed for a single mechanical resonator, it is natural to extend it to two resonators using frequency multiplexing. In our experiment we used two simultaneous readout pumps corresponding to two modulation frequencies $f_\textrm{mod,1}$ and $f_\textrm{mod,2}$. As a result, the in-phase and out-of phase quadratures of our readout contained information on both drums: $I(t)=I_1(t)+I_2(t)$ and $Q(t)=Q_1(t)+Q_2(t)$ where $I_j$ and $Q_j$ ($j\in\{1,2\}$) have the same form as in Eq.~\eqref{eq:IQofXP}. Therefore, extracting $X_1\equiv X_1(t_s)$, $X_2 \equiv X_2(t_s)$, $P_1\equiv P_1(t_s)$, $P_2 \equiv P_2(t_s)$ up to added noise, requires characterizing $\Gamma_{b,j}$, $f_{\textrm{mod},j}$ and $G_j$ for $j\in\{1,2\}$.

\begin{figure}[!h]
	\includegraphics{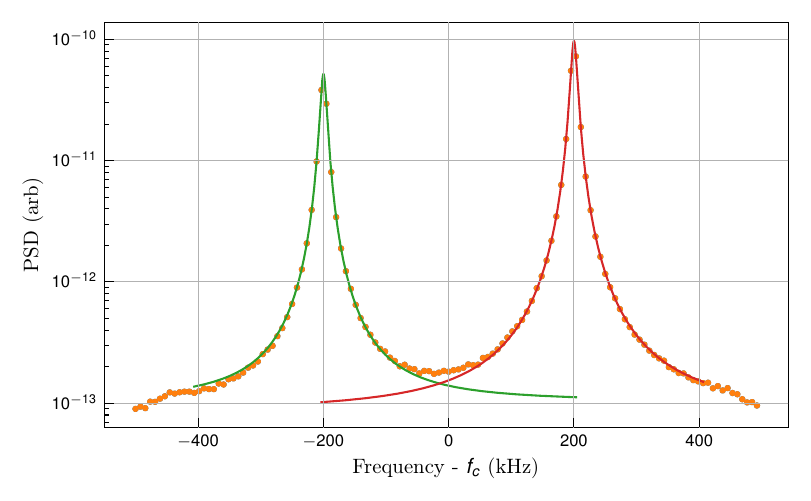}
	\caption{Power spectral density (PSD) of blue sideband readout of two drums. Solid lines are individual fits for Lorentzian functions.}\label{fig:FFT}
\end{figure}
Nominally, $f_\textrm{mod,1}=\SI{200}{\kilo\hertz}$, $f_\textrm{mod,2}=\SI{-200}{\kilo\hertz}$ and $\Gamma_{b,1}=\Gamma_{b,2}=2\pi\times \SI{66.7}{\kilo\hertz}$. We estimate the power spectral density (PSD) of our readout pulses by applying Fast Fourier Transform to the readout data. Figure~\ref{fig:FFT} shows an example of a PSD, demonstrating two distinct sidebands $\sim\SI{400}{\kilo\hertz}$ apart. We fit both sidebands to a Lorentzian in order to extract an in situ value for $f_\textrm{mod,1}$, $f_\textrm{mod,2}$ (Lorentzian centers), $\Gamma_{b,1}$ and $\Gamma_{b,2}$ ($2\pi$ times the Lorentzian full-width-half-maxima). Typically the nominal and fitted values differ by less than a percent. To account for these small changes, the fit procedure is run for every new set of  experimental sequence repetitions, i.e., if an experiment is run $N=10,000$ times, then those $N$ runs will be used to form a single PSD that would render the four fit parameters. 

To estimate the gain transduction factors $G_1$ and $G_2$, we generalize a well-used technique previously applied to single-drum devices (see for example~\mycite{Teufel2011coolng,Palomaki710,Reed2017,Delaney2019}). Briefly, we choose a desired dilution refrigerator temperature $T_{\textrm{mc}}=\SI{40}{\milli\kelvin}$, so that, on the one hand, the device mechanical parameters (frequency and quality factor) are close to those at base temperature ($\sim\SI{7}{\milli\kelvin}$) and, on the other hand, the temperature is large enough to allow for thermalization. The latter allows us to apply the equipartition theorem, i.e., to assume that the thermal occupancy $n_j$ in mechanical resonator $j$ is proportional to $\tfrac{k_bT_{\textrm{mc}}}{h f_{\textrm{m},j}}$, where $k_b$ is the Boltzmann constant and $h$ is the Planck constant. We perform an experiment where both drums are allowed to thermalize to $T_{\textrm{mc}}$ for $\SI{55}{\milli\second}$, followed by our two-mode blue-sideband readout. By applying the linear filters we estimate $\tilde{X}_j$ and $\tilde{P}_j$. The estimated $\tilde{X}_j^2$ is related to $X_j^2$ by the relation $\tilde{X}_j^2=G_j(X_j^2+n_{\textrm{add},j})=G_j(n_j+\tfrac{1}{2}+n_{\textrm{add},j})$ and similarly for $\tilde{P}_j$ . We factor $G_j$ out by equating the mechanical energy $(\tilde{X}_j^2+\tilde{P}_j^2)/2$  to $\tfrac{k_bT_{\textrm{mc}}}{h f_{\textrm{m},j}}+1$. Here we set $n_{\textrm{add},j}=\tfrac{1}{2}$, which is the minimal added noise constrained by quantum mechanics. In reality, the added noise is larger. However, at $\SI{40}{\milli\kelvin}$, the correction to $G_j$ due to a higher value of added noise is less than $5\%$, as shown below.

Our model for the measurement chain, from mechanical variables to measured variables is that of a beam-splitter: $x_j = \sqrt{\eta_j}X_j + \sqrt{1-\eta_j}\xi_j$ and similarly for $p_j$, where $\xi_j$ is a corresponding Gaussian noise operator with zero mean and $\tfrac{1}{2}$ variance. It is straightforward to show that $x_j = \tilde{X}_j/(\sqrt{G(1+2n_{\textrm{add},j})})$ and similarly for $p_j$ would satisfy the beam-splitter relation using the measurement efficiencies $\eta_j=1/(1+2n_{\textrm{add},j})$. Those are our measured variables in the correct units of $\sqrt{\textrm{quanta}}$. Recall that the measurement efficiencies were found using the theory for the variance of the drums vs. entanglement pulse duration (subsection~\ref{subsec:theory_params}). Our estimated measurement efficiencies correspond to added noise values of $n_{\textrm{add},1}=\naddone$ and $n_{\textrm{add},2}=\naddtwo$. The thermal occupancy for drums 1 and 2 respectively are $\sim 75$ and $\sim 52$ at a cryostat temperature of  $\SI{40}{mK}$. This is self consistent with less than a $5\%$ error in the original estimation of $G_j$ for $j\in\{1,2\}$. 

\begin{figure}[!h]
	\includegraphics{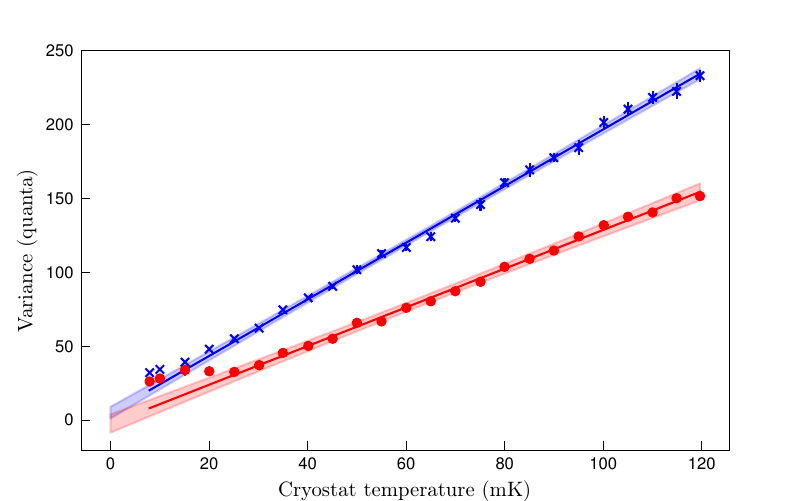}
	\caption{Drums' variances versus dilution refrigerator mixing chamber temperature. Solid lines are linear fits. Shaded regions are $1\sigma$ confidence intervals on the fits. Blue corresponds to drum 1 and red to drum 2.}\label{fig:tscan}
\end{figure}

To verify that the drums thermalize to the dilution refrigerator at $\SI{40}{\milli\kelvin}$, we vary the temperature and monitor the variance of the drums . Figure~\ref{fig:tscan} shows a linear relation between temperature and drums' variances, extending from $\SI{20}{\milli\kelvin}$ to $\SI{120}{\milli\kelvin}$. Indeed, at temperatures colder than $\SI{20}{\milli\kelvin}$ the drums start to thermally decouple from the dilution refrigerator temperature. This phenomenon has been reported previously~\mycite{Teufel2011coolng,Palomaki710,Reed2017, Delaney2019} for various temperatures at the $\SI{10}{\milli\kelvin}$ to $\SI{20}{\milli\kelvin}$ range. Our calibration point of $\SI{40}{\milli\kelvin}$  is safely within the thermalized, linear regime.

\subsection{Phase alignment}
We acquired $N=10,000$ repetitions of each experiment. Due to a limitation of our acquisition system, only $400$ repetitions could be recorded at a time. Therefore we achieved $10,000$ repetitions by accumulating $25$ chunks of data, each with $400$ repetitions. Since the local oscillator phase changed between chunks, each shot of the experiment contained a pilot tone, for absolute phase reference of the local oscillator (See table~\ref{tbl:pulses}). In addition, each drum's individual phase-space distribution was slightly displaced from the origin. The angle of the displaced thermal state was different for different chunks. In order to undo this chunk-to-chunk mismatch, we allowed an additional single-drum rotation degree of freedom per chunk and corrected for this rotation during analysis.

\subsection{Two-drum thermal decoherence}\label{subsec:decoherence}
Maintaining long coherence times is a key requirement for the entanglement experiment. Here, the coherence time of the mechanical oscillators was limited by thermal decoherence, quantified by  $\tau_\textrm{coh}$ which is the expected time it takes the oscillator to absorb a single thermal phonon (see Ref.~\mycite{Aspelmeyer:2014ce} p.1397-1398 and references therein). 

\begin{figure}[!h]
	\includegraphics{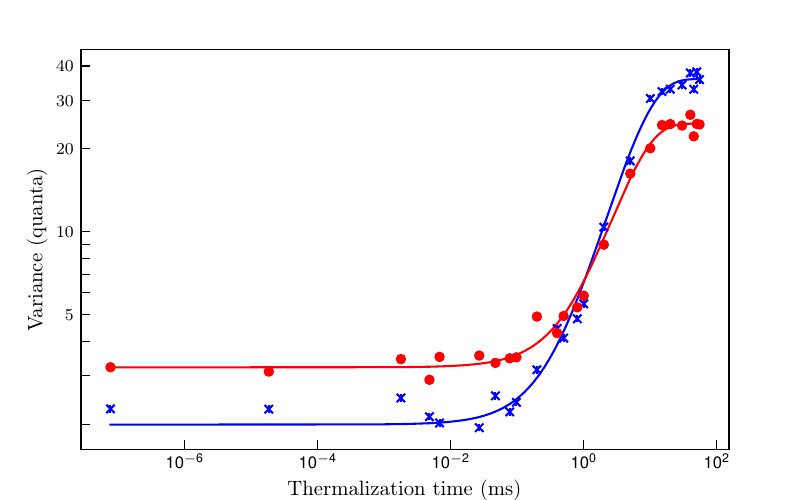}
	\caption{Thermalization experiment at a cryostat temperature of $\sim\SI{7}{\milli\kelvin}$. Blue and red circles correspond to drum 1 and drum 2 measured variances. Solid lines are fits.  }\label{fig:thermalization}
\end{figure}

We quantify the coherence times of the drums using a re-thermalization experiment. Each experimental sequence is composed of sideband cooling of the drums followed by a wait period ("thermalization time") and a two-mode readout. We plot the variances of the drums versus thermalization time in Fig.~\ref{fig:thermalization}. Both drums' variances increase by one quanta at the $\SI{100}{\micro\second}$ time scale. To better quantify this, we fit the variance versus time data to $n_\textrm{init}+n_\textrm{thermal}(1-\exp(-t/T_1))$, where $n_\textrm{init}$ is the initial cold variance, $n_\textrm{thermal}$ is the long-time thermal contribution to the variance, $t$ is the thermalization time and $T_1$ the time scale for thermalization. The fit parameters can be found in table~\ref{tbl:thermalization}. Using these parameters we estimate the thermal coherence times $\tau_\textrm{coh}= T_1/n_\textrm{thermal}$, yielding $\SI{200(10)}{\micro\second}$ and $\SI{270(20)}{\micro\second}$ for drum 1 and 2 respectively. Our entangling experiment duration was kept a factor of $\sim 10$ shorter than these time scales in order to avoid the effect of decoherence during the entangling process. 

\begin{table}[!h]
	\centering
	\caption{Fit parameters for thermalization curves} \label{tbl:thermalization}
	\begin{tabular}{cccc}
		Drum & $n_\textrm{init}$ & $n_\textrm{thermal}$ & $T_1$                        \\ \hline
		1    & 2.0(3)            & 33.9(6)              & $\SI{6.9(5)}{\milli\second}$\\
		2	& 3.2(2)			& 21.5(4)				& $\SI{5.8(5)}{\milli\second}$
	\end{tabular}
\end{table}

\section{Data analysis}\label{sec:dataanalysis}
\subsection{Estimating the measured covariance matrix}
We measure the quadratures of motion of the two mechanical oscillators at the end of each repetition of an experiment. For repetition $n$ we get the sequence $\vec{s}^{(n)} = [ x^{(n)}_{1} \hspace{1mm} p^{(n)}_{1} \hspace{1mm} x^{(n)}_{2} \hspace{1mm} p^{(n)}_{2} ] $. The estimator for the measured covariance matrix is:
\begin{equation}\label{eq:CmeasEst}
	\pazocal{C}_{\textrm{meas},i , j} = \frac{1}{N}\sum_{n=1}^N s_i^{(n)} s_j^{(n)} - \frac{1}{N^2}\sum_{n=1}^Ns_i^{(n)} \sum_{m=1}^N s_j^{(m)}.
\end{equation}
In this paper, each experiment was repeated $N=10,000$ times. 

The measured minimal symplectic eigenvalue $\numeas$ was extracted from $\pazocal{C}_\textrm{meas}$ using Eq.~\eqref{eq:numeasgen}. To assign a confidence interval (CI) to $\numeas$ we have employed non-parametric bootstrapping~\mycite{Efron:1994} where $M$ number of resampled datasets are drawn, with replacement, from the original dataset. For each one of the resampled datasets we evaluate the corresponding witness $\numeas^{(m)}$ where $m=1 \ldots M$. Finally the CI is assigned through the bias-corrected percentile method~\mycite{Buckland:1984}.  In this work we used $M=1,000$. 


\subsection{Estimating the covariance matrix before loss}
To estimate the covariance matrix before the loss we solve the following Semi-Definite Program (SDP) implemented in the python convex optimization package CVXPY:

\begin{align}\label{eq:sdp}
	\underset{\pazocal{C}}{\textrm{minimize}}& \Big\vert \Big\vert \hspace{1mm} \pazocal{C}_{\text{meas}} - \pazocal{F}[\pazocal{C}] \hspace{1mm} \Big\vert \Big\vert _{2},\\
	\textrm{subject to\hspace{3mm}}&  \pazocal{C} + i \frac{\Omega} {2} \geq 0,\label{eq:Heisenberg}
\end{align}

\noindent where $\pazocal{C}_{\text{meas}}$ is the covariance matrix after  loss (Eq.~\eqref{eq:CmeasEst}), $\pazocal{C}$ is the covariance matrix of the mechanical oscillators, $ \Omega = \left(\begin{smallmatrix}
	J&\mathbf{0}\\\mathbf{0}&J
\end{smallmatrix}\right)$, $J = \left(\begin{smallmatrix}0 &1 \\ -1 & 0 \end{smallmatrix}\right) $ and $\mathbf{0}=\left(\begin{smallmatrix}0 &0 \\ 0 & 0 \end{smallmatrix}\right) $. The function $\pazocal{F}[\pazocal{C}]$ mixes the covariance matrix with vacuum through beam-splitters (Eq.~\eqref{eq:beamsplitter}) and traces out the environmental modes. The constraint in Eq.~\eqref{eq:Heisenberg} imposes the Heisenberg uncertainty principle on each one of the mechanical modes~\mycite{Simon2000}. This guarantees that $\pazocal{C}$ is a physically-realizable covariance matrix of two harmonic oscillators. In many cases, $\pazocal{F}^{-1}(\pazocal{C}_\textrm{meas})$ already satisfies Eq.~\eqref{eq:Heisenberg} and the optimization becomes trivial: $\pazocal{C}=\pazocal{F}^{-1}(\pazocal{C}_\textrm{meas})$ regardless of the choice of norm (here $l_2$) in Eq.~\eqref{eq:sdp}. In the case where $\pazocal{F}^{-1}(\pazocal{C}_\textrm{meas})$ is not physically realizable, the optimization is actually required. Once $\pazocal{C}$ is obtained, $\numech$ can be calculated using Eq.~\eqref{eq:numech}. We obtain statistical confidence intervals on $\numech$ using non-parametric bootstrapping, similar to the method used for $\numeas$. We checked that the dependence of the estimation process on our choice of $l_2$-norm in Eq.~\eqref{eq:sdp} is negligible compared to the CI. Specifically, we ran the optimization process for the $l_1$, $l_2$, $l_\infty$ norms as well as the nuclear norm (sum of all singular values). The resulting differences in $\numech$ were at the $1\times 10^{-5}$ relative error or less, which is indeed negligible compared to the CI.  

\subsection{Systematic uncertainty}\label{subsec:systematics}
Estimating $\numeas$ and $\nu$ requires knowing the measurement efficiencies $\eta_1$ and $\eta_2$, obtained independently, and the readout amplification rate, measured in situ (subsection~\ref{subsec:from_raw_data_to_units_of_quanta}). Therefore, our uncertainty on the values of $\eta_1=\etaone$, $\eta_2=\etatwo$ and $\Gamma_{b,j}\pm\Delta\Gamma_{b,j}$ translate into systematic error on $\numech$ and $\numeas$.  The approach for systematic estimation is identical for both cases, and we describe it in terms of $\numech$ for brevity. To quantify systematic uncertainty, we compute three values for $\nu$ using three pairs of values for $\eta_1$ and $\eta_2$: $\nu_L$ uses $\eta_1=0.24$ and $\eta_2=0.150$, $\nu_C$ uses $\eta_1=0.26$ and $\eta_2=0.153$, and $\nu_R$ uses $\eta_1=0.28$ and $\eta_2=0.156$.  Noting that $\nu_L < \nu_C < \nu_R$, we calculate systematic errors $\delta\nu_{\eta,+}=\nu_R-\nu_C$ and $\delta\nu_{\eta,-}=\nu_C-\nu_L$.  Similarly, we calculate $\delta\nu_{\textrm{gain,}\pm}$ by calculating $\nu$ with values of the gain that correspond to $\Gamma_{b,j}\pm\Delta\Gamma_{b,j}$. We combine these systematic errors incoherently $\sigma_{\textrm{sys},\pm}=(\delta\nu_{\textrm{eta,}\pm}^2+\delta\nu_{\textrm{gain,}\pm}^2)^{1/2}$. The center value $\numech_C$ with its statistical confidence interval is plotted in Fig.~3D in black circles. The inset of Fig.~3D shows $\numech_C$ (black circle) as well as $\numech_C+\sigma_{\textrm{sys},+}$, $\numech_C-\sigma_{\textrm{sys},-}$ (triangles).  Figure~3C follows the same conventions.

\subsection{Blind analysis of the data}
We tested and fine-tuned our data analysis procedure on numerically simulated data as well as on a measured data set ("training set"). The fit that we described in subsection~\ref{subsec:theory_params} used the training set. Based on the monotonicity of the fit, we decided that the last point acquired in the scan of entanglement versus pulse duration ($\SI{16.8}{\micro\second}$) would be used to report the entanglement bound achieved. 

The data analysis protocol was applied to a fresh data set  ("published set") without any modifications of the protocol obtained with the training set. This is the data that appears in Fig.~3 of the paper. The paper also reports the entanglement achieved at $\SI{16.8}{\micro\second}$ entanglement pulse duration, as decided prior to the analysis of the published set. 

\appendix
\section{Experimental parameters}
For convenience we summarize the experimental system parameters in Table~\ref{tbl:parameters}.

\begin{table}[!h]
	\caption{Measured and estimated experimental parameters at a cryostat temperature of $\sim\SI{7}{\milli\kelvin}$. Mechanical thermal coherence times are defined in Sec.~\ref{subsec:decoherence}.} \label{tbl:parameters}
	\begin{tabular}{lll}
		Parameter&					Symbol&			Value\\
		\hline
		Mechanical resonance frequency (drum 1)& $f_\textrm{m,1}$&  $\SI{10.865}{\mega\hertz}$\\
		Mechanical resonance frequency (drum 2)& $f_\textrm{m,2}$&  $\SI{15.898}{\mega\hertz}$\\
		Mechanical $1/e$ decay time (drum 1)&						$T_{1,1}$	&	$\SI{6.9}{\milli\second}$\\
		Mechanical $1/e$ decay time (drum 2)&						$T_{1,2}$	&	$\SI{5.8}{\milli\second}$\\
		Mechanical thermal coherence time (drum 1)&		$\tau_{\textrm{coh},1}$	&	$\SI{200}{\micro\second}$\\
		Mechanical thermal coherence time (drum 2)&		$\tau_{\textrm{coh},2}$	&	$\SI{270}{\micro\second}$\\
		Microwave cavity resonance frequency&	$\fc$&	$\SI{6.0806}{\giga\hertz}$\\
		Microwave cavity total linewidth&	$\kappa$&	$2\pi\times \SI{800}{\kilo\hertz}$\\
		Bare electromechanical coupling (drum 1)&	$g_{0,b}$& $2\pi\times\SI{17}{\hertz}$\\
		Bare electromechanical coupling (drum 2)&	$g_{0,r}$& $2\pi\times\SI{22}{\hertz}$\\
		Electromechanical coupling (drum 1)& $\Gamma_b$& $2\pi \times\SI{82}{\kilo\hertz}$\\
		Electromechanical coupling (drum 2)& $\Gamma_r$& $2\pi \times\SI{94}{\kilo\hertz}$\\
		Readout efficiency (drum 1)& $\eta_1$& $26\%$\\
		Readout efficiency (drum 2)& $\eta_2$& $15.3\%$
	\end{tabular}
\end{table}

\renewcommand{\refname}{SUPPLEMENTARY REFERENCES}

\end{document}